\documentclass[preprint,amsmath,showpacs,superscriptaddress]{revtex4}
\usepackage{graphicx}
\usepackage{epstopdf}
\usepackage{bm}
\usepackage{epsfig}
\usepackage{graphics}
\usepackage{xspace}
\usepackage{amsfonts}
\usepackage{comment}

\def\Slash#1{{#1\!\!\!\slash}}

\newcommand{\nn}{\nonumber}

\newcommand{\bea}{\begin{eqnarray}}
\newcommand{\eea}{\end{eqnarray}}

\begin{document}


\title{Searching for the signal of dark matter and photon associated production at the LHC beyond leading order}

\vspace*{1cm}

\author{Fa Peng Huang}
\affiliation{School of Physics and State Key
Laboratory of Nuclear Physics and Technology, Peking
University, Beijing, 100871, China}

\author{Chong Sheng Li\footnote{Electronic
address: csli@pku.edu.cn}}
\affiliation{School of Physics and State Key
Laboratory of Nuclear Physics and Technology, Peking
University, Beijing, 100871, China}
\affiliation{Center for High Energy Physics, Peking
University, Beijing, 100871, China}

\author{Jian Wang}
\affiliation{School  of Physics and State Key
Laboratory of Nuclear Physics and Technology, Peking
University, Beijing, 100871, China}

\author{Ding Yu Shao}
\affiliation{School of Physics and State Key
Laboratory of Nuclear Physics and Technology, Peking
University, Beijing, 100871, China}



\begin{abstract}
 \vspace*{0.3cm}

We study the signal of dark matter and photon associated production
induced by the vector and axial-vector operators at the LHC,
including the QCD next-to-leading order (NLO) effects.
We find that the QCD NLO corrections reduce the dependence of the total
cross sections on the factorization and renormalization scales,
and the $K$ factors increase with  the increasing of the dark matter mass,
which can be as large as about 1.3
for both the vector and axial-vector operators.  Using our QCD NLO results,
we improve the constraints on the new physics scale from the results
of the recent CMS experiment. Moreover, we  show the Monte Carlo simulation results
for detecting the $\gamma+\Slash{E}_{T}$ signal at the QCD  NLO level,
and  present the integrated luminosity needed for a $5\sigma$ discovery
at the $14$ TeV LHC  . If the signal is not observed,
the lower limit on the new physics scale can be set.
\end{abstract}


\maketitle
\newpage

\section{Introduction}
\label{sec:1}
The dark matter (DM)  attracts a lot of attention in the fields of both cosmology and
particle physics  \cite{{Jungman:1995df, Bertone:2004pz}}. The astrophysical
observations have provided strong evidence for the existence of DM \cite{Komatsu:2010fb}.
Compared to the direct and indirect experiments, the hadron colliders have impressive
advantages that the measurements are not sensitive to the
uncertainties related to the galactic distributions, DM velocities, etc.
There have been a lot of studies to search for DM at the LHC
in a series of DM models \cite{Buchmueller:2011ki,
Profumo:2011zj,Belanger:2011ny,Kile:2011mn,Akula:2011dd,Feldman:2011me,
Bai:2010hd,Gogoladze:2010ch,Cheung:2010zf,Goodman:2010ku,Bertone:2010rv,
Giudice:2010wb,Li:2010rb,Beltran:2010ww,Zhang:2009dd,ArkaniHamed:2008qp,
Fargion:1995qb}.
We can probe the DM through the visible particles, which are associated
produced, such as a photon or a jet \cite{Bai:2010hh, Fox:2011pm, Haisch:2012kf}.
In this work, we only consider the DM and photon associated production at the LHC,
since this signal is clear and suffers from less backgrounds from the standard model (SM).

Recently, the CMS collaboration has searched for new physics (NP) in the $\gamma+\Slash{E}_{T}$ final state,
and set the $90\%$  confidence level  (C.L.) lower limits on the NP scale for
vector and axial-vector operators  \cite{Chatrchyan:2012tea}.
However, the  analysis for the DM searching there is based on the leading order (LO) results, which suffer
from large uncertainties due to the choice of renormalization and factorization scales.
In our previous work \cite{Wang:2011sx}, we only considered the QCD next-to-leading order (NLO) corrections  for the
case of the scalar operator. Following the ideas of our previous works \cite{Wang:2011sx,Gao:2009pn}, in this paper,
we study the signal of DM and photon associated production induced by  the
vector and axial-vector operators at the LHC, including  QCD NLO  corrections.
Using our NLO results, we improve the constraints on the NP scale from the
results of  recent CMS experiment.

In Sec.~\ref{sec:operator}, we show the vector and axial-vector operators
describing  the interactions between DM and the SM particles.
In Sec.~\ref{sec:relic}, we show the constraints on  the DM  mass and the
NP scale from the relic abundance.
In Sec.~\ref{sec:nlo}, the numerical results are presented and discussed.
In Sec.~\ref{sec:background}, we
analyze the backgrounds in  the SM and discuss the discovery potential
at the $14$ TeV LHC.
Section~\ref{sec:conclusion}  contains a brief conclusion.
\section{Vector and Axial-Vector Operators}
\label{sec:operator}
We consider the dimension six vector and axial-vector operators
\begin{eqnarray}    \label{eq-operator}
  \mathcal{O}_{V}&=&\frac{\kappa}{\Lambda^{2}}(\bar{q}\gamma^{\mu}q)(\bar{\chi}\gamma_{\mu}\chi),   \nonumber  \\
  \mathcal{O}_{A}&=&\frac{\kappa}{\Lambda^{2}}(\bar{q}\gamma^{\mu}\gamma_{5}q)(\bar{\chi}\gamma_{\mu}\gamma_{5}\chi),
\end{eqnarray}
which are also studied in Refs. \cite{Beltran:2008xg, Cao:2009uw, Goodman:2010ku, Bai:2010hh, Mambrini:2011pw}.
The NP scale $\Lambda$ can be regarded as the remnant of integrating the
massive propagator between the DM and SM particles.
We assume that the Dirac fermion $\chi$ is a DM candidate,
and  a singlet under the SM gauge group $SU(3)_{c}\times SU(2)_{L} \times U(1)_{Y}$.
The DM $\chi$ can only interact with the quarks by these operators.
In Ref. \cite{Chatrchyan:2012tea}, the  constraints on the NP scale $\Lambda$ for the vector and axial-vector
operators are given by the CMS collaboration through the process of DM and photon associated production at LO.
In this paper, we will perform the QCD NLO corrections to these processes, whose effects are important
for research at the LHC,  and improve the limits on the NP scale.

\section{Constraints from the Relic Abundance}
\label{sec:relic}

\begin{figure}
  \includegraphics[width=0.6\linewidth]{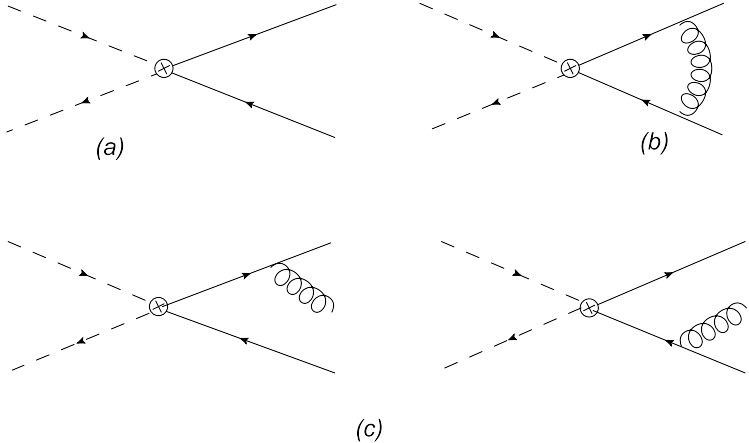}\\
  \caption{Feynman diagrams for the DM annihilation process. Labels (a), (b)
and (c) denote the LO, virtual correction and real corrections, respectively.}
  \label{fig-annihilation}
\end{figure}

Before discussing the signal of the DM at the LHC, we first consider the
relic abundance  which is a precise observable in cosmology.
The relic abundance can impose strong constraints on the properties of the DM,
and can be obtained  from the DM annihilation cross section.
The Feynman diagrams are shown in Fig. \ref{fig-annihilation}.
First, we get the LO  annihilation cross section for the vector operator
\begin{eqnarray} \label{myvborn}
  \sigma_{B,V}^{\rm{an}}  =N_c N_f \frac{\kappa^2}{\Lambda^4}\sqrt{\frac{s}{s-4m^2}} \frac{s+2m^2}{12\pi},
\end{eqnarray}
where $s$ is the square of center-of-mass energy.
$N_c$ and $N_f$ are the numbers of  color and flavor of quarks, respectively. $m$ is the mass
of the DM. This LO  cross section in Eq. (\ref{myvborn}) is consistent with the unexpanded result
in Eq. (8) of Ref. \cite{Beltran:2008xg}.
For the case of vector operator, we get the QCD NLO corrections
\begin{equation}
   \sigma^{an}_{\rm{NLO},V}=K^{\rm{an}}\sigma_{B,V}^{\rm{an}},
\end{equation}
where  $K^{\rm{an}}=1+\alpha_s/ \pi$ is the $K$ factor of the cross section, generally defined as $\sigma_{\rm{NLO}}/\sigma_{\rm{LO}}$.

We assume that the DM  is moving at nonrelativistic velocities ($v \ll 1$) when freezing out.
We define $v$ as the relative velocity between the DM so that the square of the center-of-mass energy can
be written as $s \approx 4m^2+m^2v^2+m^2 v^4/4$. Thus we can expand
\begin{equation}
  \sigma^{an}_{\rm{NLO},V} v \approx a + b v^2,
\end{equation}
where
\begin{eqnarray} \label{eqvexpand}
  a &=& K^{\rm{an}}N_c N_f\frac{\kappa^2}{\Lambda^4}\frac{m^2}{\pi}, \nn \\
  b &=& K^{\rm{an}}N_c N_f\frac{\kappa^2}{\Lambda^4}\frac{m^2}{6\pi}.
\end{eqnarray}
For the case of axial-vector operator, we follow the same process and give only the main results.
The LO cross section is
\begin{eqnarray}
  \sigma_{B,A}^{\rm{an}}  =N_c N_f \frac{\kappa^2}{\Lambda^4}\sqrt{\frac{s}{s-4m^2}} \frac{s-4m^2}{12\pi},
\end{eqnarray}
which is  consistent with the unexpanded result in Eq. (9) of Ref. \cite{Beltran:2008xg}.
After including the QCD NLO corrections to the process induced by the axial-vector operator, we get
\begin{equation}
  \sigma^{\rm{an}}_{\rm{NLO},A} v \approx a + b v^2,
\end{equation}
where
\begin{eqnarray} \label{eqaexpand}
  a &=& 0, \nn \\
  b &=& (1+\frac{\alpha_s }{\pi})N_c N_f\frac{\kappa^2}{\Lambda^4}\frac{m^2}{6\pi}.
\end{eqnarray}
Our expanded results in Eqs. (\ref{eqvexpand}) and (\ref{eqaexpand}) are consistent with the results in Refs. \cite{Bai:2010hh,Fox:2011fx}.

\begin{figure}
  \includegraphics[width=0.49\linewidth]{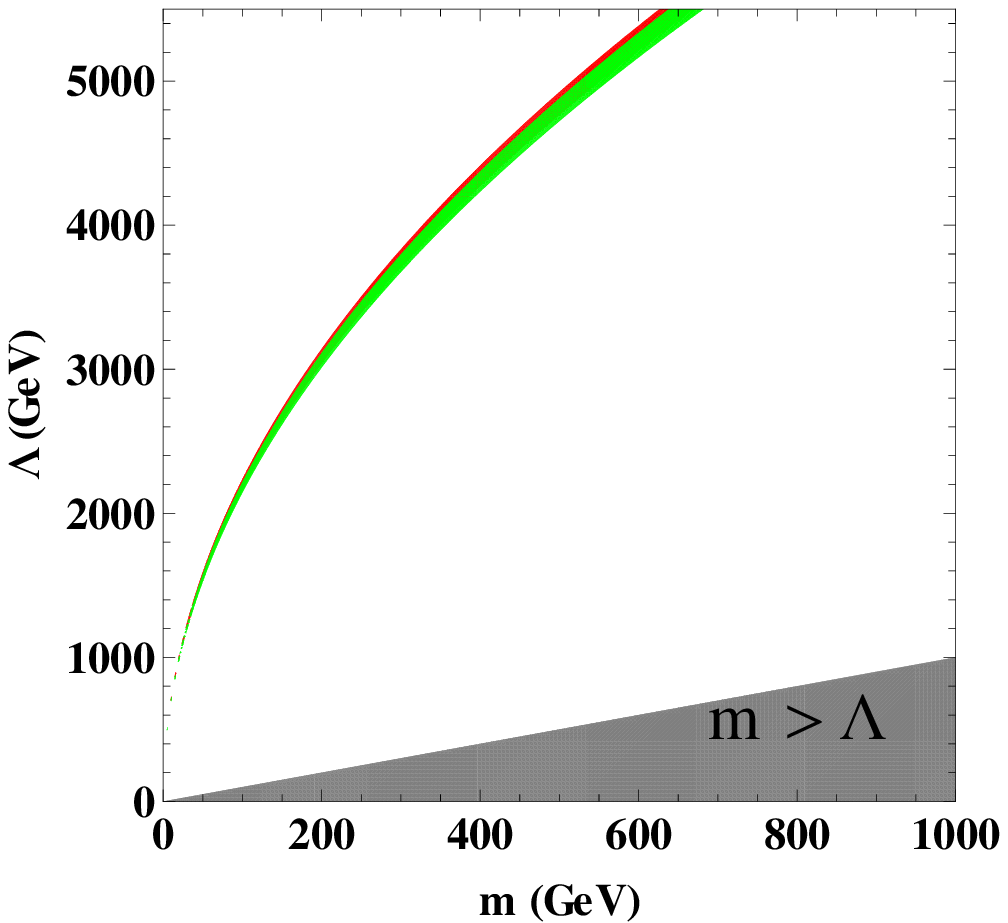}
  \includegraphics[width=0.49\linewidth]{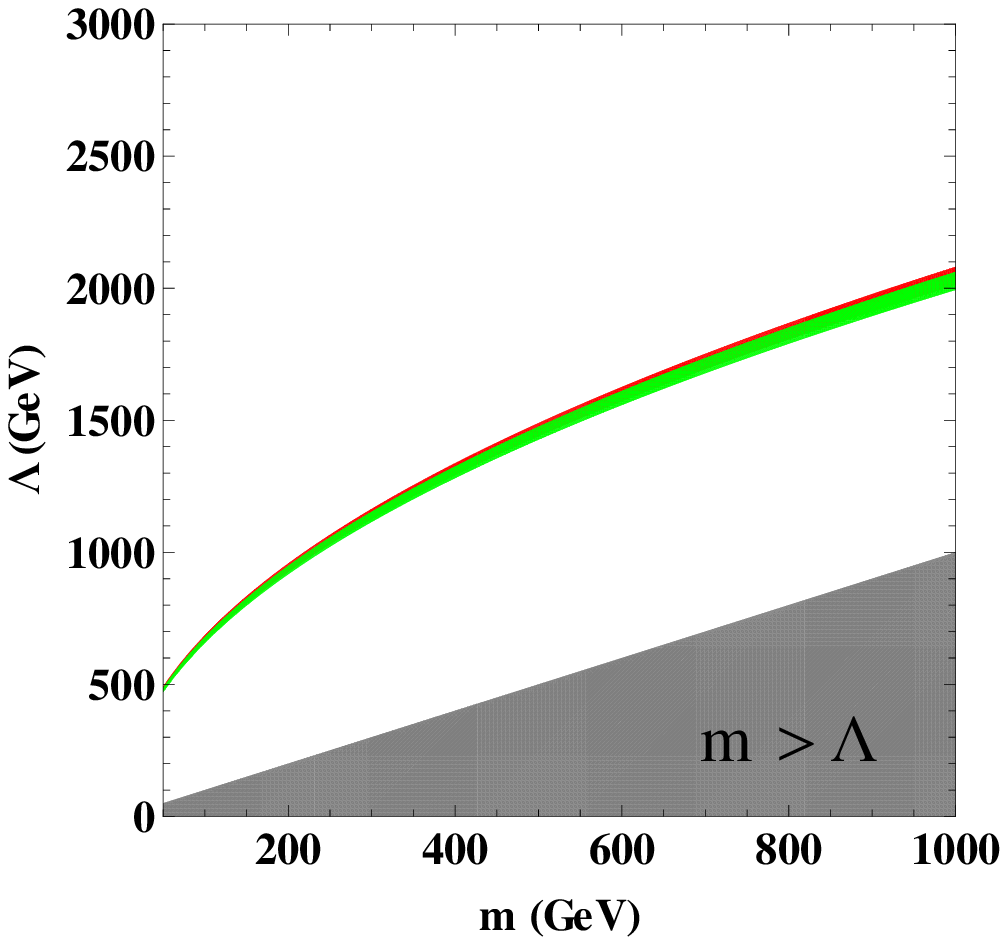}\\
  \caption{ Constraints on  the DM  mass and the
NP scale for the vector (left) and axial-vector (right) operators, respectively .
The relic abundance is required to be in the $2\sigma$ region
around the observed central value \cite{Jarosik:2010iu}. The lower green band is the LO result. The upper red band is
the NLO result. In this figure, we choose $\kappa=1, \alpha_s=0.118$ and $ N_f=5$. }
  \label{fig-relic}
\end{figure}

We perform the calculation of relic abundance by using the method in Ref. \cite{Kolb:1990vq}.
Then we show the constraints on the NP scale
and DM mass in Fig. \ref{fig-relic}.
We see that the NLO corrections increase the lower limits on the NP scale slightly.
The regions below the red band are allowed, since we assume that this kind of DM is not the unique candidate.
\section{Numerical Results of the QCD  NLO CORRECTIONS for the Case of the Vector and Axial-Vector Operators}
\label{sec:nlo}
Different from the scalar operator in our previous work \cite{Wang:2011sx}, the quark sector and the DM
sector can not factorize for the vector and axial-vector operators that we consider in this paper. This leads to more complicated
analytical expressions in our calculation.
We follow the same approach in our
previous paper \cite{Wang:2011sx} to cancel the infrared (IR) divergences in QCD NLO calculations,
and  show the numerical results for the case of  vector and axial-vector operators  below.

\subsection{QCD corrections for the case of the vector and the axial-vector operators}

\begin{figure}
  \includegraphics[width=0.6\linewidth]{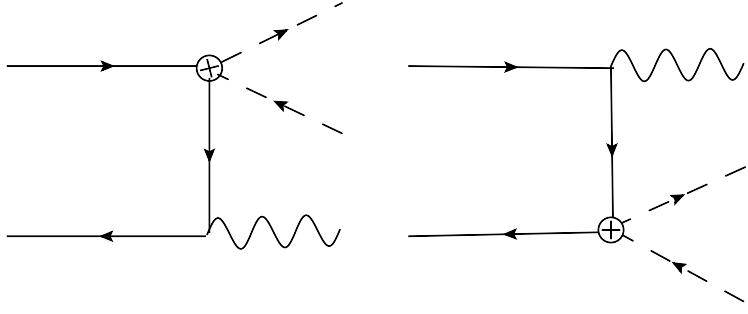}\\
  \caption{LO Feynman diagrams for the DM and photon associated production.}
  \label{fig-born}
\end{figure}

First of all, we calculate the LO cross section of the following process
\begin{equation}\label{eq-process}
    q(p_1)+\bar{q}(p_2) \to \chi(p_3)+\bar{\chi}(p_4)+\gamma(p_5).
\end{equation}
The LO Feynman diagrams are shown in Fig. \ref{fig-born}.
The LO partonic cross section is
\begin{equation}\label{eq-LOpartonic}
    \hat{\sigma}_B=\frac{1}{2s_{12}}\int d \Gamma_3 \overline{|\mathcal{M}_B|^2},
\end{equation}
where $\Gamma_3$ is the three-particle final states phase space.
We define $s_{ij}=(p_i+p_j)^2$, $t_{ij}=(p_i-p_j)^2$ and $\alpha=e^2/4\pi$.
The spin and color summed and averaged Born matrix element squared  is
\begin{eqnarray}
  \overline{|\mathcal{M}_B|^2} &=&
  \frac{\pi\alpha \kappa^2}{3\Lambda^4} \sum_{i} Q_{i}^{2}  |\mathcal{M}_0^{V(A)}|^2,
\end{eqnarray}
where $|\mathcal{M}_0^{V}|^2$ for the case of the vector operator is expressed as
\begin{eqnarray}
    |\mathcal{M}_0^{V}|^2 &=&
    \frac {1} {t_ {15} t_ {25}} 16 (2 m^4 (4 s_ {12} +
      t_ {15} + t_ {25} ) +  m^2 (-s_ {12} (2 s_ {35}                                \nonumber \\ &&
      + 2 s_ {45} + 3 t_ {13} + 3 t_ {14} - 4 t_ {15} +
          3 t_ {23} + 3 t_ {24} - 4 t_ {25} )                                               \nonumber \\ &&
          - s_ {45} t_ {15} - s_ {45} t_ {25} -
      s_ {35} (t_ {15} + t_ {25} ) +                                                     \nonumber  \\ &&
      4 (s_ {12} ) {}^2 + 2 (t_ {15} ) {}^2 +
      2 (t_ {25} ) {}^2 - 2 t_ {13} t_ {15} -
      2 t_ {14} t_ {15} + t_ {15} t_ {23} + t_ {15} t_ {24} +   \nonumber  \\ &&
      t_ {13} t_ {25} + t_ {14} t_ {25} - 2 t_ {23} t_ {25} -
      2 t_ {24} t_ {25} ) + s_ {45} t_ {13} t_ {15} +           \nonumber   \\ &&
   s_ {35} t_ {14} t_ {15} +
   s_ {12} (s_ {45} (t_ {13} + t_ {23} ) +                      \nonumber   \\ &&
      s_ {35} (t_ {14} + t_ {24} ) +
      2 (t_ {14} t_ {23} + t_ {13} t_ {24} ) ) +
   s_ {45} t_ {23} t_ {25} + s_ {35} t_ {24} t_ {25} +        \nonumber     \\ &&
   t_ {14} t_ {15} t_ {23} + t_ {13} t_ {15} t_ {24} -
   2 t_ {15} t_ {23} t_ {24} - 2 t_ {13} t_ {14} t_ {25} +    \nonumber     \\ &&
   t_ {14} t_ {23} t_ {25} + t_ {13} t_ {24} t_ {25} ),
\end{eqnarray}
and $Q_{i} (i=1,5)$ are the electric charge of the  quarks.
For the case of  axial-vector operator,
\begin{equation}
|\mathcal{M}_0^{A}|^2=|\mathcal{M}_0^{V}|^2-\frac{64 m^2\left(2 s_ {12}\left(t_ {15} + t_ {25} \right) +2\left(s_ {12}\right){}^2 + \left (t_ {15}\right) {}^2+\left(t_ {25} \right){}^2 \right)}{t_ {15} t_ {25}}.
\end{equation}

The LO total cross section at the LHC is obtained by convoluting the partonic
cross section with the parton distribution functions (PDFs) $G_{q(\bar{q})}(x)$ :
\begin{equation}\label{eq-LOhadronic}
    \sigma_B=\int d x_1 d x_2 [G_{q/p}(x_1)G_{\bar{q}/p}(x_2)+(x_1 \leftrightarrow x_2)]\hat{\sigma}_B.
\end{equation}

The QCD NLO  corrections consist of real gluon radiation, quark or
antiquark emission and one-loop virtual gluon effects. We use dimensional
regularization to regulate both the ultraviolet (UV) and the IR divergences
in our calculations.

After renormalization, the UV divergences in the virtual
corrections are removed,  leaving the IR divergences and the finite
terms.
\begin{figure}
  \includegraphics[width=0.6\linewidth]{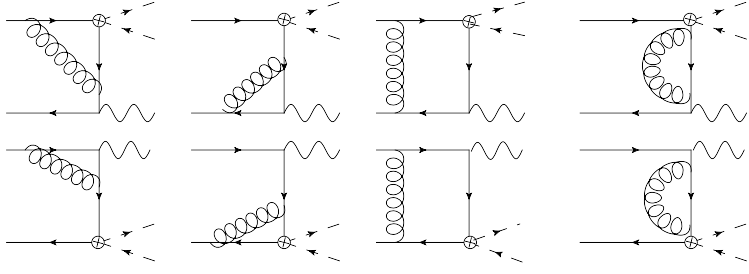}\\
  \caption{Feynman diagrams for one-loop virtual corrections.}
  \label{fig-virt}
\end{figure}
The final virtual gluon corrections to the partonic cross section are
\begin{equation}\label{eq-VirtPartonic}
    \hat{\sigma}_v=\frac{1}{2s_{12}}\int d \Gamma_3 2\overline{Re(\mathcal{M}_B^{*}\mathcal{M}_v)},
\end{equation}
for which the Feynman diagrams are shown in Fig. \ref{fig-virt}.
The IR divergent part of $\mathcal{M}_{v}$ is given by
\begin{equation}
 \mathcal{M}_{v}^{IR}=\frac{\alpha_s }{4\pi}C_{\epsilon}\left(\frac{A_2^v}{\epsilon^2}+\frac{A_1^v}{\epsilon^1}\right)\mathcal{M}_B,
\end{equation}
where $C_{\epsilon}=\Gamma(1+\epsilon)[(4\pi\mu_R^2)/s_{12}]^\epsilon$ and
\begin{eqnarray}
  A_2^v &=& -2C_F, \nn\\
  A_1^v &=& -3C_F.
\end{eqnarray}

\begin{figure}
  \includegraphics[width=0.6\linewidth]{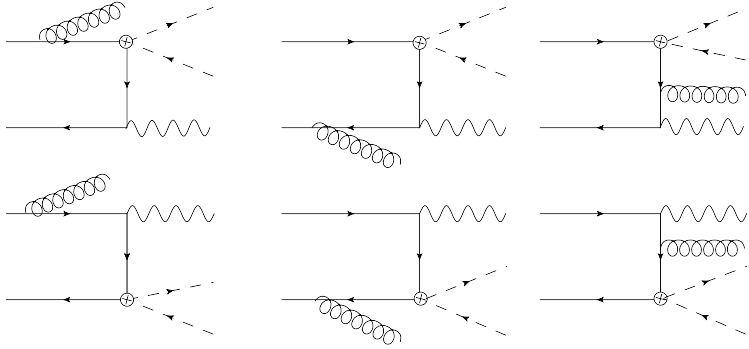}\\
  \caption{Feynman diagrams for a real gluon emission.}
  \label{fig-real}
\end{figure}

The Feynman diagrams for the real gluon radiation process
\begin{equation}\label{eq-GluonEmission}
    q(p_1)+\bar{q}(p_2) \to \chi(p_3)+\bar{\chi}(p_4)+\gamma(p_5)+g(p_6)
\end{equation}
are shown in Fig. \ref{fig-real}. Soft and collinear divergences appear when
we perform the final state phase integrations. To cancel the IR singularities,
we use the two cutoff phase space slicing method to integrate the singular
regions analytically \cite{Harris:2001sx}.  Explicitly, we use the soft cutoff parameter
$\delta_s$ to define the soft regions and the collinear cutoff parameter $\delta_c$ to define
the hard collinear regions. The soft regions are just the phase space where the real radiated
gluon's energy $E_6\le \delta_s\sqrt{s_{12}}/2$. The collinear regions are defined by  $|t_{i6}|<\delta_c s_{12}$ with $i=1,2$.
Thus, the partonic cross section of the real gluon radiation can be separated as
\begin{equation}\label{eq-Real}
    d\hat{\sigma}_r=d\hat{\sigma}_r^S+d\hat{\sigma}_r^{HC}+d\hat{\sigma}_r^{\overline{HC}}.
\end{equation}
Here, $\hat{\sigma}_r^S$ and $\hat{\sigma}_r^{HC}$ represent the partonic
cross section for the soft regions and hard collinear regions, respectively.
The hard noncollinear part $\hat{\sigma}_r^{\overline{HC}}$ is finite and
can be computed numerically using standard Monte Carlo integration techniques.

In the soft regions, using the eikonal approximation, the cross section can be factorized as
\begin{equation}\label{eq-Soft2}
    d\hat{\sigma}_r^S=d\hat{\sigma}_B\frac{\alpha_s }{2\pi}C_{\epsilon}
    \left(\frac{A_2^S}{\epsilon^2}+\frac{A_1^S}{\epsilon}+A_0^S\right),
\end{equation}
where
\begin{equation}
    A_2^S=2C_F,\qquad A_1^S=-4C_F\ln \delta_s.
\end{equation}

In the hard collinear limits, the squared matrix element factors
into the product of a splitting kernel and  a leading order squared
matrix element. Then we obtain
\begin{eqnarray}\label{eq-HChadronic}
  d\sigma_r^{HC} &=& d\hat{\sigma}_B \frac{\alpha_s}{2\pi}C_{\epsilon}
  \left(-\frac{1}{\epsilon}\right)\delta_c^{-\epsilon}[P_{qq}(z,\epsilon)G_{q/p}(x_1/z)
  G_{\bar{q}/p}(x_2) \nn \\
  &+& P_{\bar{q}\bar{q}}(z,\epsilon)G_{\bar{q}/p}(x_1)
  G_{q/p}(x_2/z)  + (x_1 \leftrightarrow x_2)]
  \frac{dz}{z}\left( \frac{1-z}{z} \right)^{-\epsilon}dx_1 dx_2,
\end{eqnarray}
in which the $P_{ij}(z,\epsilon)$ are the unregulated splitting function.
To factorize the collinear singularity into the PDFs, we use scale dependent
PDFs in the $\overline{\rm MS}$ convention:
\begin{equation}\label{eq-Redifinepdf}
    G_{b/p}(x,\mu_F)=G_{b/p}(x)+\left(-\frac{1}{\epsilon}\right)
    \left[\frac{\alpha_s}{2\pi}\frac{\Gamma(1-\epsilon)}{\Gamma(1-2\epsilon)}
    \left(\frac{4\pi\mu_R^2}{\mu_F^2}\right)^{\epsilon}\right]\int_x^1\frac{dz}{z}
    P_{ba}(z)G_{a/p}(x/z).
\end{equation}
Now, we replace $G_{q(\bar{q})/p}$ in the LO hadronic cross section (\ref{eq-LOhadronic})
and combine the result with the hard collinear contribution in Eq. (\ref{eq-HChadronic}). The
resulting $\mathcal{O}(\alpha_s)$ expression for the initial state collinear contribution is
\begin{eqnarray}
  d\sigma^{coll} &=& d\hat{\sigma}_B\frac{\alpha_s}{2\pi}C_{\epsilon}
  \Big\{ \tilde{G}_{q/p}(x_1,\mu_F)G_{\bar{q}/p}(x_2,\mu_F)+
  G_{q/p}(x_1,\mu_F)\tilde{G}_{\bar{q}/p}(x_2,\mu_F) \nn \\
  &+& \sum_{a=q,\bar{q}}\Big[ \frac{A_1^{sc}(a\to a g)}{\epsilon} + A_0^{sc}(a\to a g)\Big]
  G_{q/p}(x_1,\mu_F)G_{\bar{q}/p}(x_2,\mu_F)  \nn \\
  &+& (x_1 \leftrightarrow x_2) \Big\} dx_1 dx_2.
\end{eqnarray}
with
\begin{equation}
  A_1^{sc}(q\to q g) = C_F(2\ln \delta_s +3/2 ).
\end{equation}
The $\tilde{G}$ functions are given by
\begin{equation}
    \tilde{G}_{b/p}(x,\mu_F)=\sum_{a}\int_x^{1-\delta_s\delta_{ab}}
    \frac{dy}{y} G_{a/p}(x/y,\mu_F)\tilde{P}_{ba}(y)
\end{equation}
with
\begin{equation}
    \tilde{P}_{ba}(y)=P_{ba}(y)\ln \Big( \delta_c \frac{1-y}{y} \frac{s_{12}}{\mu_F^2} \Big)
    -P^{'}_{ba}(y).
\end{equation}

\begin{figure}
  \includegraphics[width=0.6\linewidth]{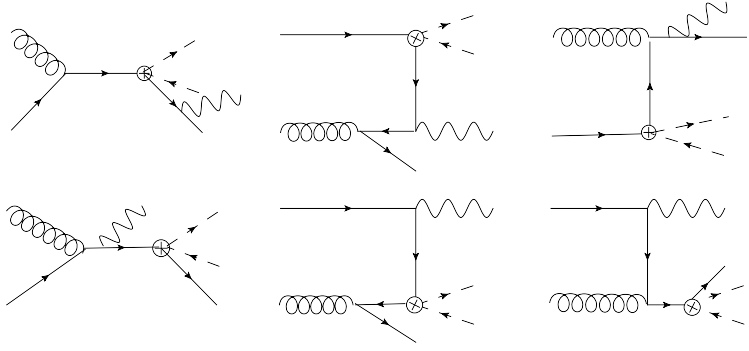}\\
  \caption{Feynman diagrams for a quark  emission.
  The Feynman diagrams for antiquark emission can be obtained by charge conjugation.}
  \label{fig-real_qg}
\end{figure}

A complete real correction includes also the (anti)quark emitted processes, as shown in Fig. \ref{fig-real_qg},
such as
\begin{eqnarray}\label{eq-QuarkEmission}
    g(p_1)+q/\bar{q}(p_2) \to \chi(p_3)+\bar{\chi}(p_4)+\gamma(p_5)+q/\bar{q}(p_6).
\end{eqnarray}
We only need to deal with the collinear
divergences which can be totally absorbed into the redefinition of the PDFs in Eq.
(\ref{eq-Redifinepdf}) for these processes.

Finally, the NLO total cross section for the process $pp\to \chi\bar{\chi}\gamma$ is
\begin{eqnarray}
  \sigma^{NLO} &=& \int dx_1 dx_2 \Big\{ \big[ G_{q/p}(x_1,\mu_F)G_{\bar{q}/p}(x_2,\mu_F)
  +(x_1 \leftrightarrow x_2) \big](\hat{\sigma}_B+\hat{\sigma}_v+\hat{\sigma}^S_r
  +\hat{\sigma}_r^{\overline{HC}}) \Big\}+\sigma^{coll} \nn \\
  &+& \sum_{a=q,\bar{q}}\int dx_1 dx_2 \big[ G_{g/p}(x_1,\mu_F)G_{a/p}(x_2,\mu_F)
  +(x_1 \leftrightarrow x_2) \big]\hat{\sigma_r}^{\overline{C}}
  (ga\to \chi \bar{\chi}\gamma a),
\end{eqnarray}
where $\overline{C}$ in $\hat{\sigma_r}^{\overline{C}}(ga\to \chi \bar{\chi}\gamma a)$ means that
the phase space integration is performed in the noncollinear regions.
Note that the above expression contains no singularities since
\begin{equation}
    A_2^v+A_2^S=0,\qquad A_1^v+A_1^S+ 2A_1^{sc}(q\to q g)=0,
\end{equation}
and we can perform numerical integration now.

\subsection{Numerical results}
We use the CTEQ6L1 (CTEQ6M) PDF sets \cite{Pumplin:2002vw} and
the corresponding strong coupling $\alpha_s$ for the LO (NLO) calculations.
The default factorization scale $\mu_F$ and renormalization scale $\mu_R$ are set as $2m$.
Recently, the observations of the gamma ray in Fermi-LAT give  the hints of $130$ GeV DM \cite{Bringmann:2012vr,Weniger:2012tx}.
Thus, we set the default parameters  $(m,\Lambda)=(130{\rm ~GeV},500{\rm ~GeV})$  and $\kappa=1$ unless otherwise specified,
which are allowed by the constraints of relic abundance. Here, we choose the kinematic cuts
\begin{eqnarray}
  p_T^{\gamma} &>& 100{\rm ~GeV},  \nn \\
  |\eta^{\gamma}| &<& 2.4,  \nn \\
  p_T^{miss} &>& 100{\rm ~GeV},  \nn \\
  p_T^{jet} &>& 20 {\rm ~GeV},      \nn \\
  |\eta^{jet}| &<& 2.5,              \nn \\
  \sum_{R_{j\gamma}\in R_0} p_T^{jet} &<& p_T^{\gamma}\Big(\frac{1-\cos R_{j\gamma}}{1-\cos R_0}\Big),
\end{eqnarray}
where $R\equiv \sqrt{\Delta \phi^2+\Delta \eta^2}$ and $R_0=0.4$.

\begin{figure}
  \includegraphics[width=0.48\linewidth]{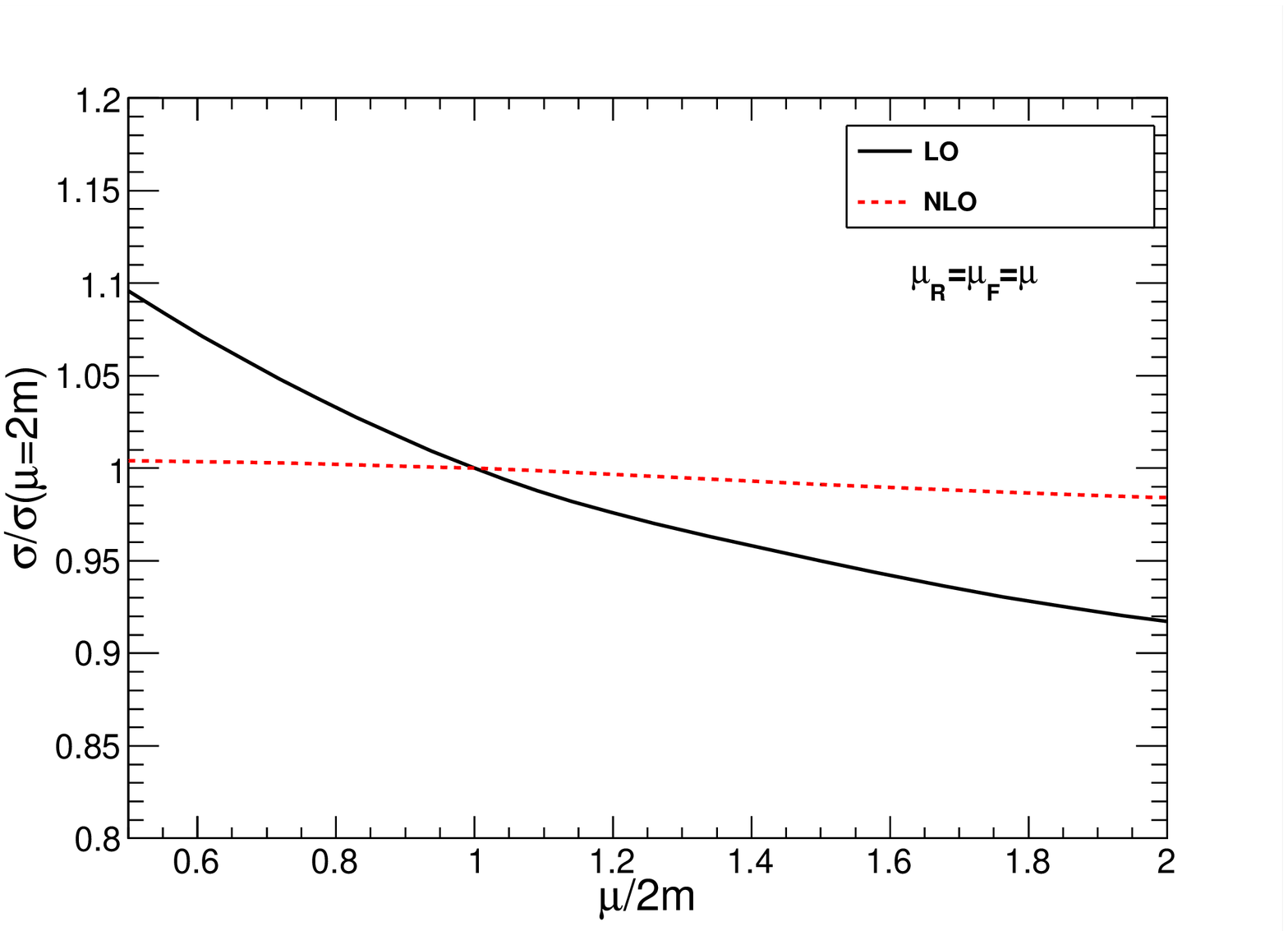}
  \includegraphics[width=0.48\linewidth]{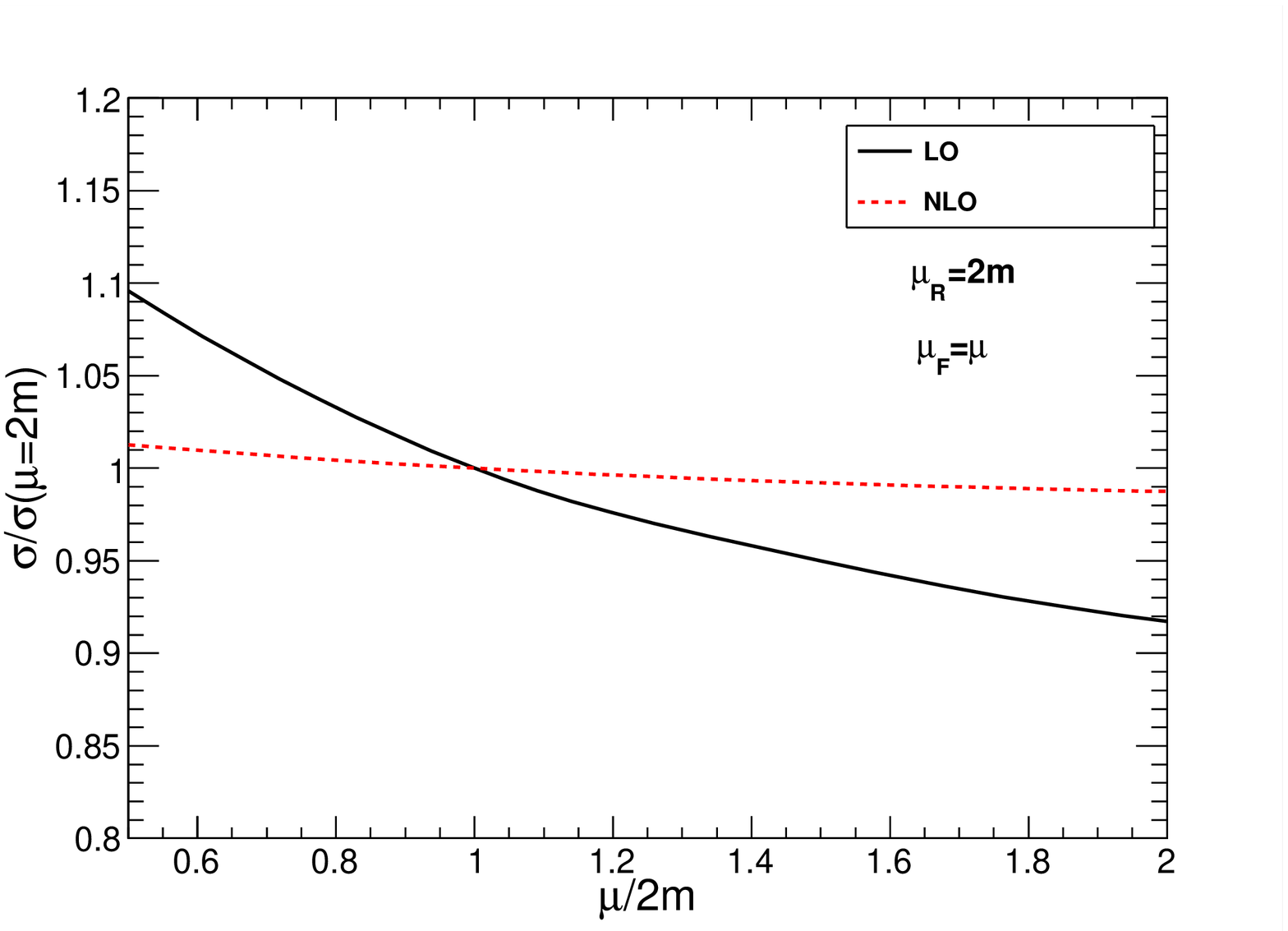}\\
  \caption{Dependence of the LO (NLO) cross sections on the factorization scale $\mu_{F}$ and
  renormalization scale $\mu_{R}$ for the vector operator.}
  \label{fig-FacScale}
\end{figure}

\begin{figure}
  \includegraphics[width=0.48\linewidth]{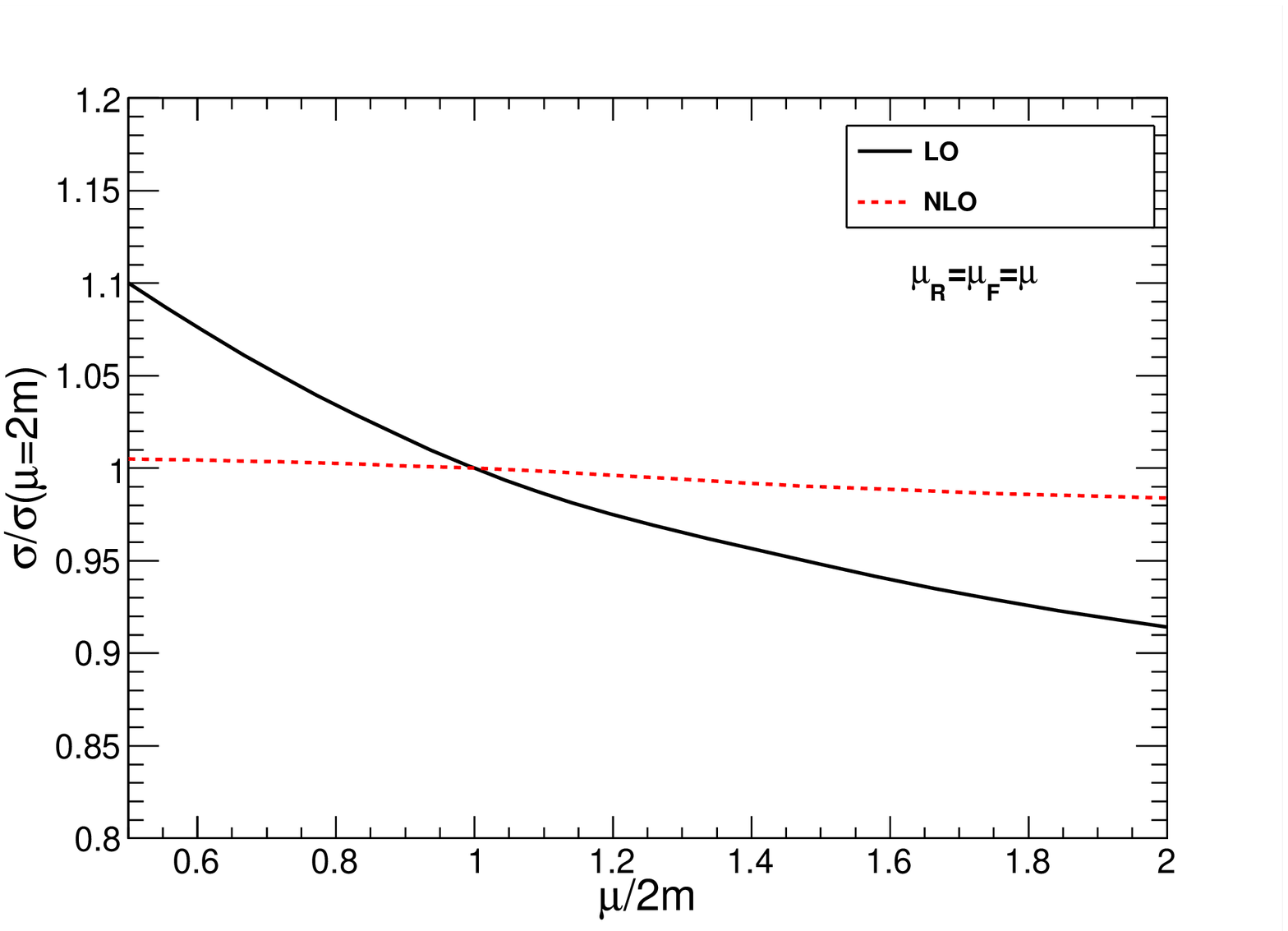}
  \includegraphics[width=0.48\linewidth]{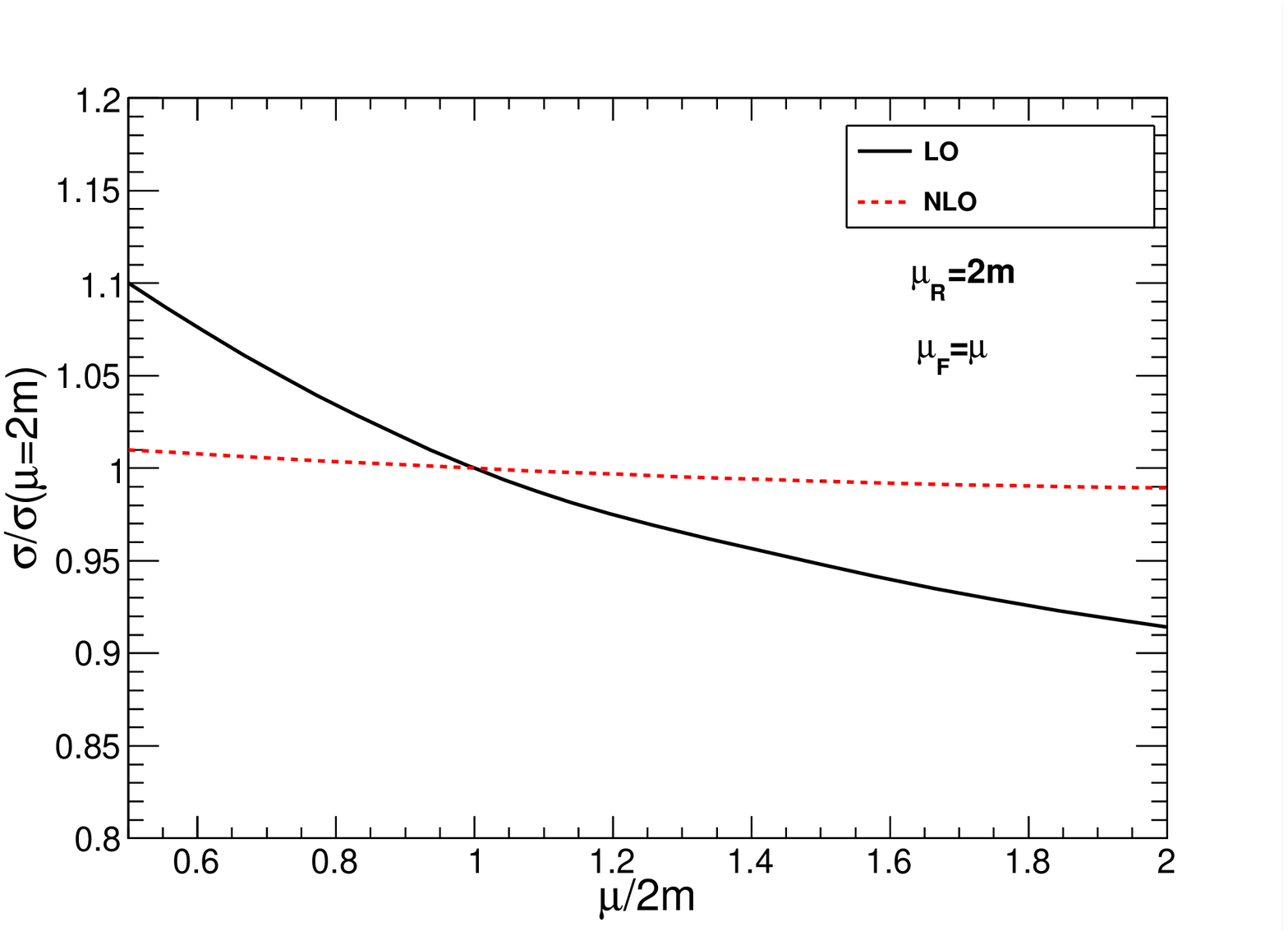}
  \caption{Dependence of the LO (NLO) cross sections on the factorization scale $\mu_{F}$ and
  renormalization scale $\mu_{R}$ for the axial-vector operator.}
  \label{fig-avFacScale}
\end{figure}

In Figs. \ref{fig-FacScale} and  \ref{fig-avFacScale}, we show the dependence of the LO (NLO) cross sections for the DM and photon
  associated production at the LHC on the factorization scale $\mu_{F}$ and
  renormalization scale $\mu_{R}$.
It can be seen that  the dependence of the NLO cross section on the factorization scale $\mu_F$
and renormalization  scale $\mu_R$ is significantly reduced, compared to the LO cross section.
This makes the theoretical prediction much more reliable.

\begin{figure}
  \includegraphics[width=0.49\linewidth]{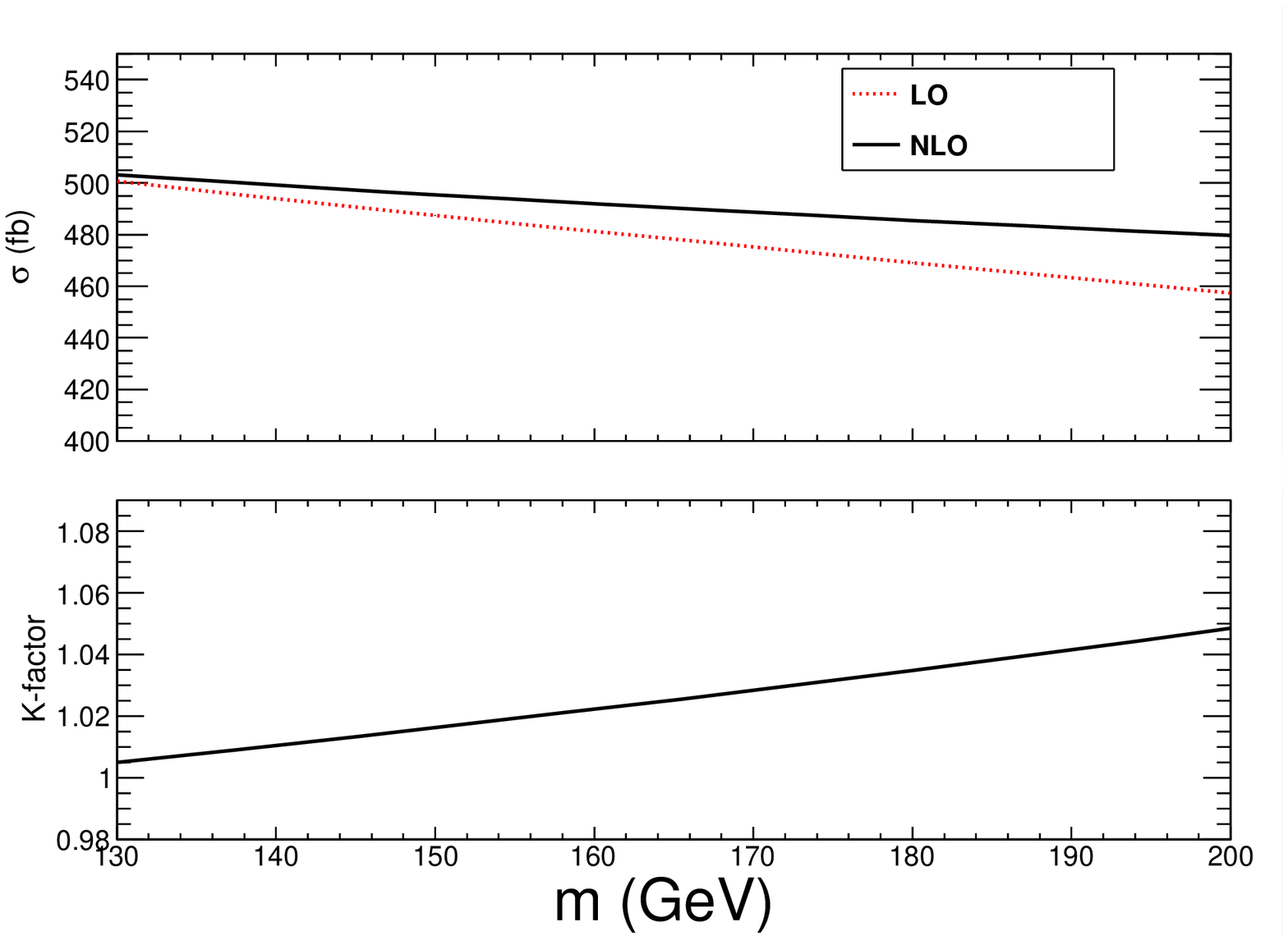}
  \includegraphics[width=0.49\linewidth]{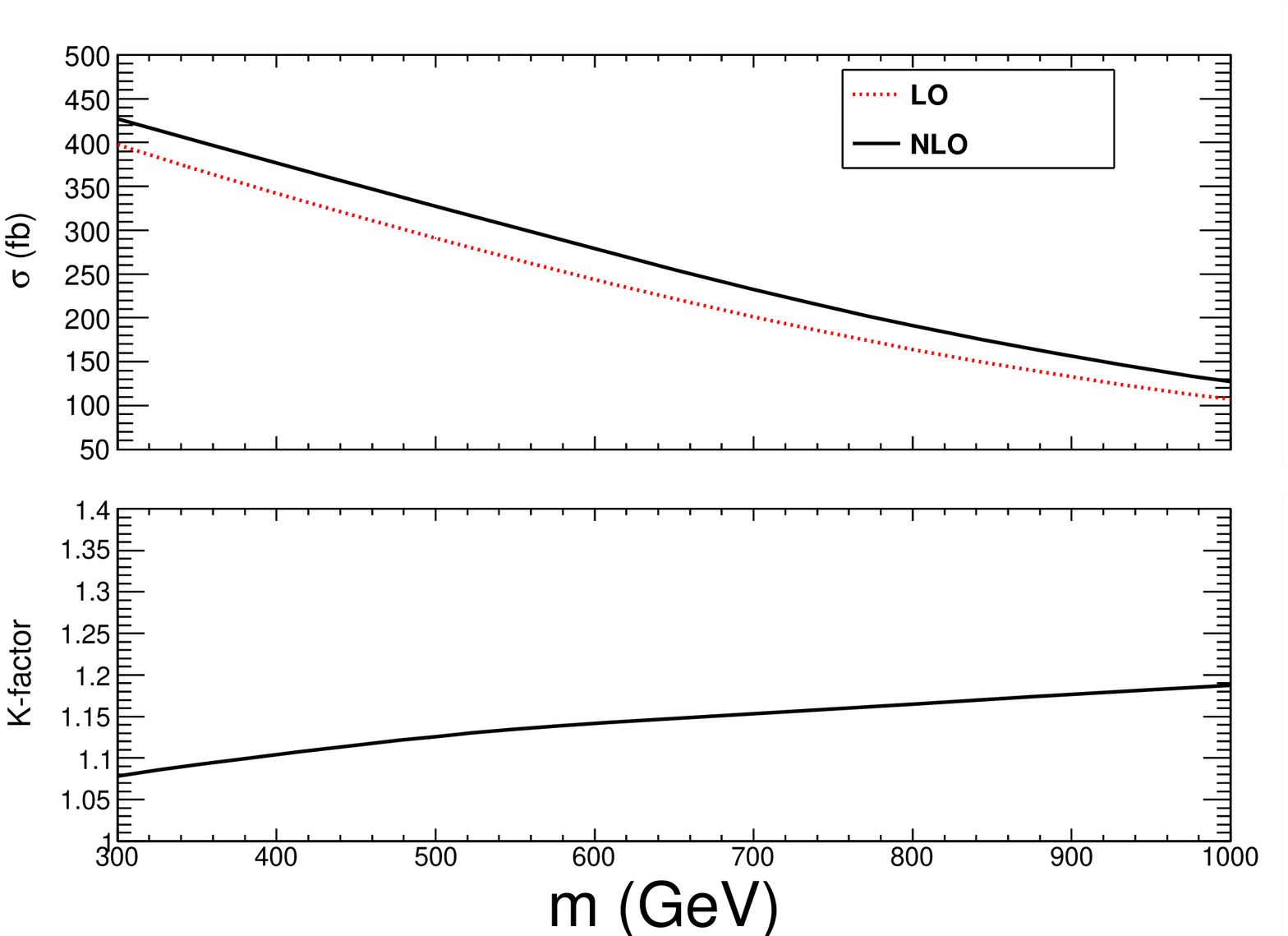}\\
  \caption{Dependence of the LO and NLO cross sections on the DM mass for the
  vector operator at $14$ TeV LHC. The $K$ factors are also shown.}
  \label{mdependence}
\end{figure}

\begin{figure}[h!]
  \includegraphics[width=0.49\linewidth]{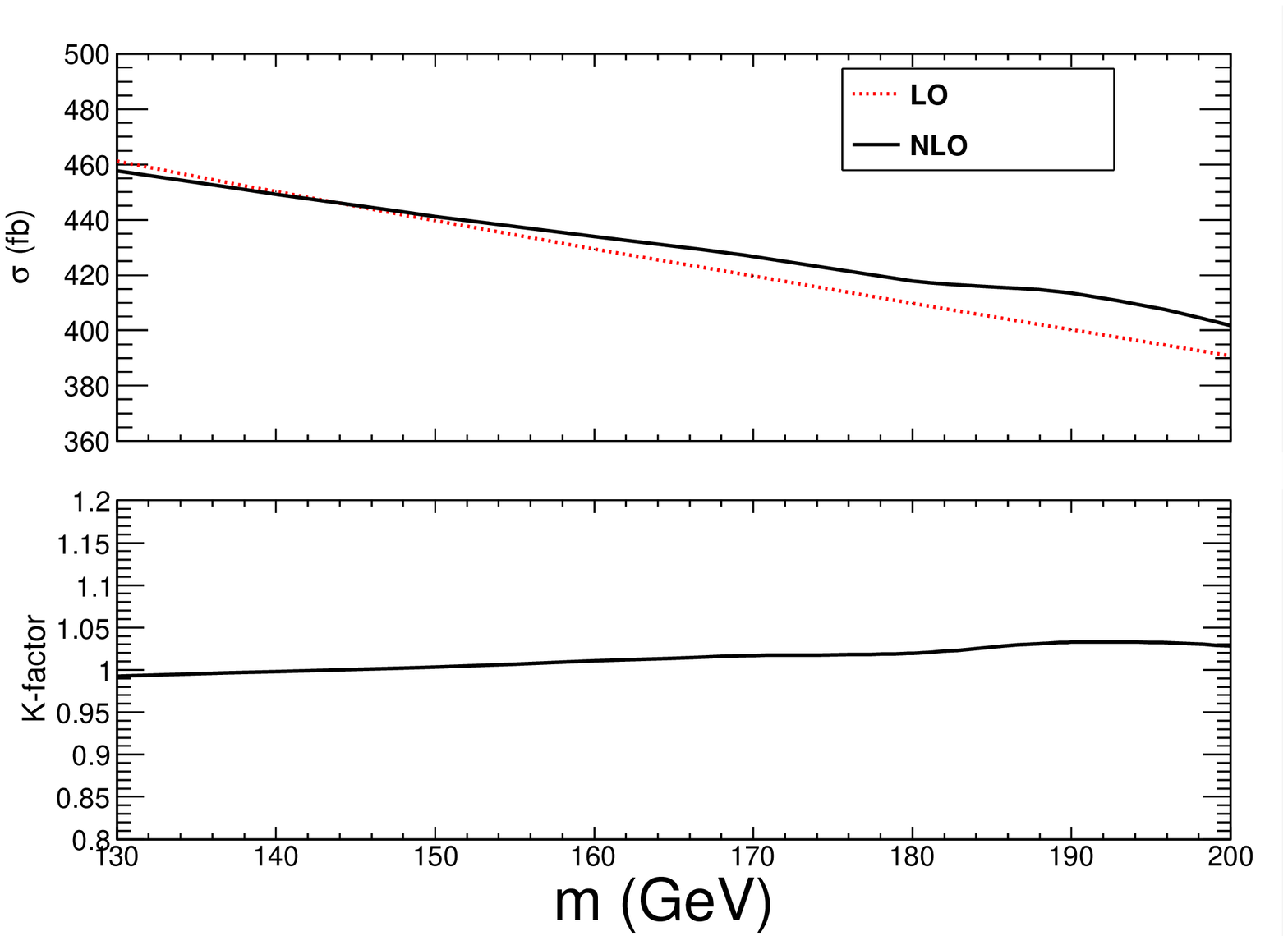}
  \includegraphics[width=0.49\linewidth]{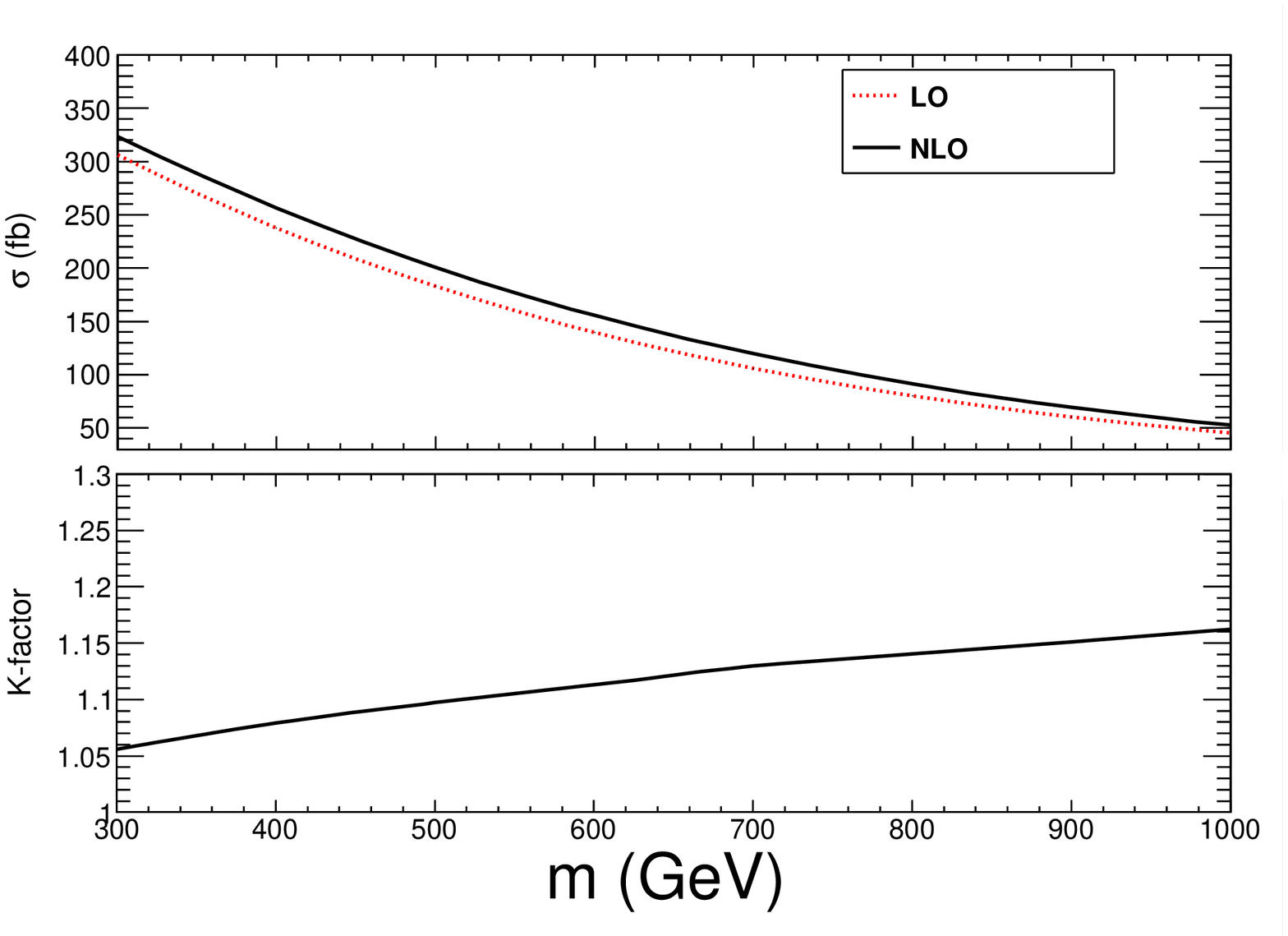}\\
  \caption{Dependence of the LO and NLO cross sections on the DM mass for the axial-vector operator
  at the $14$ TeV LHC. The $K$ factors are also shown.}
  \label{avmdependence}
\end{figure}

\begin{figure}
  \includegraphics[width=0.49\linewidth]{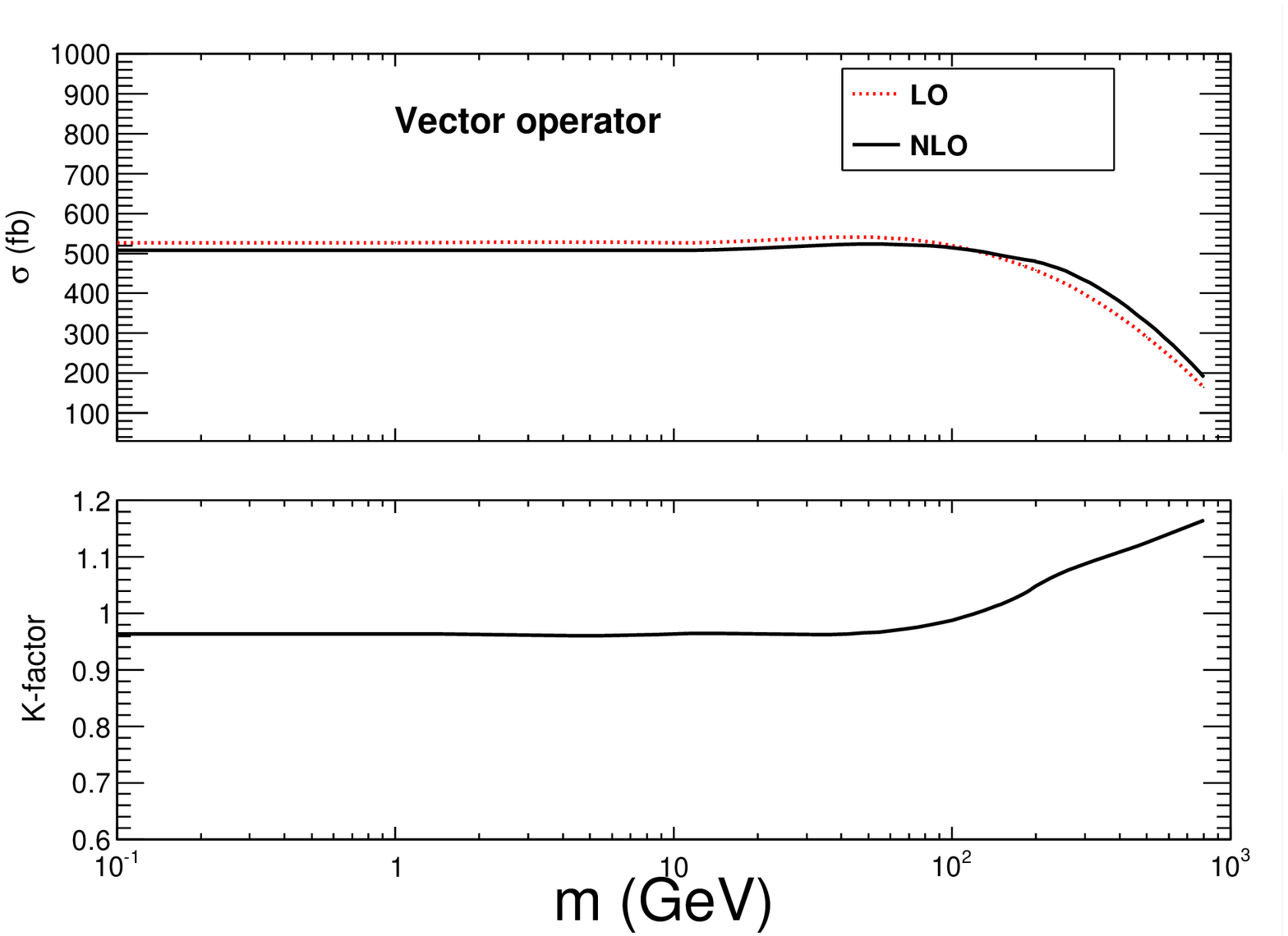}
  \includegraphics[width=0.49\linewidth]{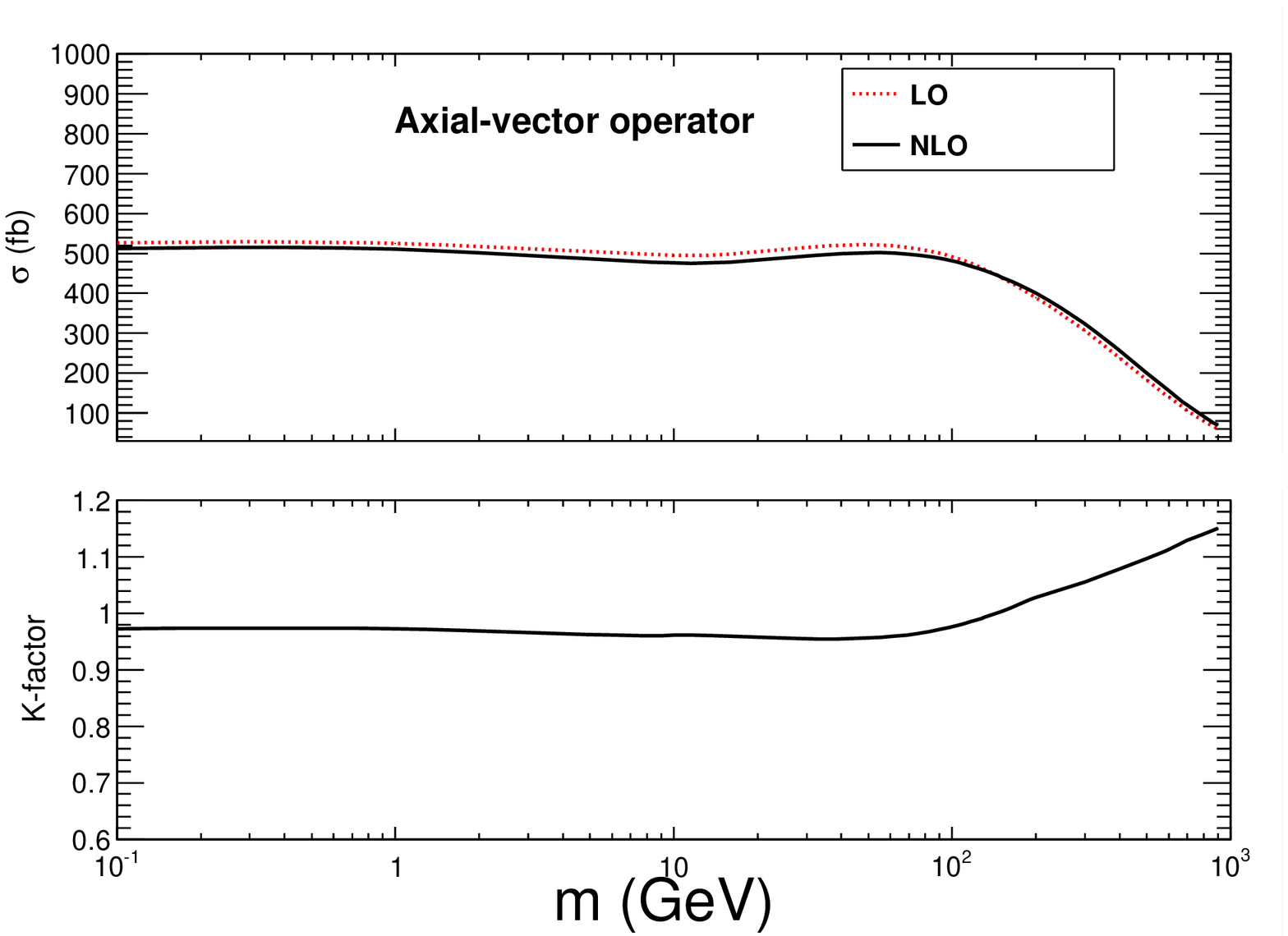}\\
  \caption{Dependence of the LO cross section, NLO cross section  and $K$ factors
  on the DM mass at the $14$ TeV LHC.}
  \label{avtot}
\end{figure}

In Fig. \ref{mdependence}, we show the DM mass dependence of the LO and NLO cross
sections  for producing heavy DM  at the $14$ TeV LHC induced by  the vector operator.
When the DM mass varies from $130$ to $200$ GeV,
the  QCD NLO corrections are modest. For the DM mass  in the range from
$300$  to $1000$ GeV, the  QCD NLO corrections generally improve the cross section
and are more significant for larger DM mass. For example,  the QCD NLO corrections  increase the cross sections by
about $19\%$ for $m=1000$ GeV.  Thus, it is
necessary to consider the NLO corrections to the process of  DM  production at hadron colliders.
We also show the mass dependence of the $K$-factors for the axial-vector operator in Fig. \ref{avmdependence},
which is similar to the case of vector operator.
Since there is no explicit limit on the DM mass, for completeness,
we show the NLO results on DM mass from $0.1$ GeV to $1000$ GeV in Fig. \ref{avtot}.
It can be seen that the $K$ factor in the light DM region, i.e. less than 100 GeV,
is nearly a constant, which is about 0.96 and 0.98 for the case of vector and axial-vector operators, respectively.

In order to compare with the experimental results of CMS, we use the
same kinematic cuts and center-of-mass energy as in \cite{Chatrchyan:2012tea},
 and improve the lower limits on the NP scale
in the results of  the  CMS  collaboration \cite{Chatrchyan:2012tea},
using our $K$ factors at the $7$ TeV LHC.
Here, we show  the improved limits on the NP scale $\Lambda$ in Table \ref{cutoff} and  \ref{avcutoff}
for the vector and axial-vector operators, respectively.

\begin{table}[h!]
\centering
\caption{Sample results of the 90\% C.L. lower limits on the NP scale $\Lambda$ for the vector operator.
The LO results are given in the CMS analysis \cite{Chatrchyan:2012tea}. The $K$ factors at
the $7$ TeV LHC  for different DM masses are also shown.}
\begin{tabular}{ccccc}
\hline
\hline
 $m$ ~[GeV]  & $\Lambda$~[GeV](LO) \cite{Chatrchyan:2012tea} &$\Lambda$~[GeV](NLO) & $K$ factor@7 TeV\\
\hline
200       & 549    & 564  & 1.11  \\
500       & 442    & 463  & 1.20 \\
1000      & 246    & 263 & 1.31 \\
\hline
\hline
\end{tabular}

\label{cutoff}
\end{table}
\begin{table}[h!]
\centering
\caption{Sample results of the 90\% C.L. lower limits on the NP scale $\Lambda$ for the axial-vector operator. The LO results
are given in the CMS analysis \cite{Chatrchyan:2012tea}. The $K$ factors at
the $7$ TeV LHC  for different DM masses are also shown.}
\begin{tabular}{ccccc}
\hline
\hline
 $m$ ~[GeV]  & $\Lambda$~[GeV](LO) \cite{Chatrchyan:2012tea} &$\Lambda$~[GeV](NLO) & $K$ factor@7 TeV\\
\hline
200       & 508    & 517  & 1.07  \\
500       & 358    & 372  & 1.17 \\
1000      & 172    & 183 & 1.29 \\
\hline
\hline
\end{tabular}

\label{avcutoff}
\end{table}

\section{Discovery Potential}
\label{sec:background}

In this section, we present the Monte Carlo simulation results for detecting the $\gamma+\Slash{E}_{T}$ signal
at the $14$ TeV LHC with NLO accuracy in perturbative QCD.
The main irreducible SM backgrounds are the $pp \to Z(\to \nu\bar{\nu})+\gamma$
and $pp \to Z(\to \nu\bar{\nu})+j$ when the jet is misidentified as a photon.
The misidentified probability is set to be $P_{\gamma/j}=10^{-4}$, as pointed out in  Ref. \cite{Baur:1992cd}.
We use the Monte Carlo program MCFM \cite{Campbell:2011bn, Baur:1997kz, Ohnemus:1992jn, Giele:1993dj}
 to calculate the backgrounds at the NLO level.

\begin{figure}
  \includegraphics[width=0.6\linewidth]{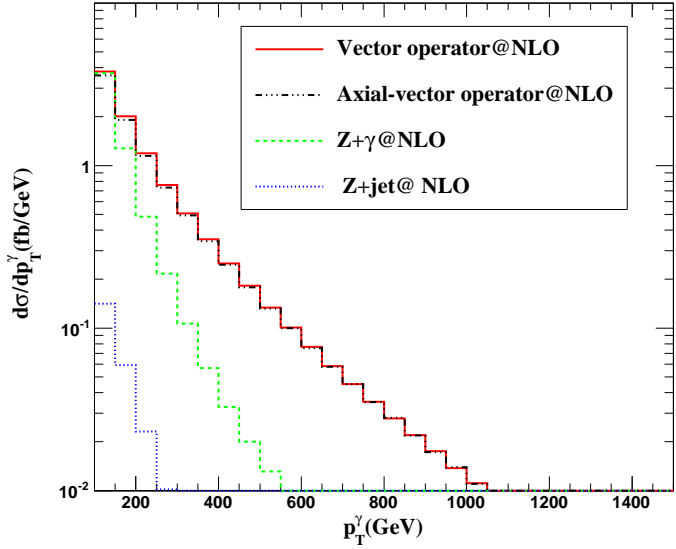}
  \caption{Dependence of the differential cross section on $p_T^{\gamma}$.}
  \label{fig-pt_a}
\end{figure}

\begin{figure}
  \includegraphics[width=0.6\linewidth]{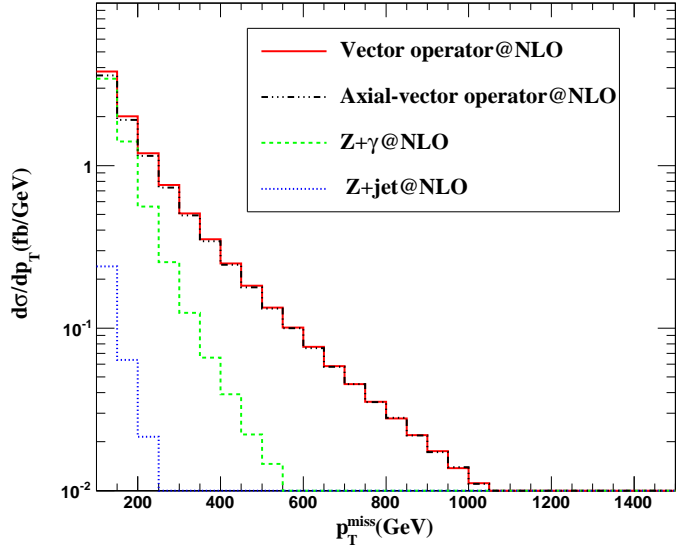}
  \caption{Dependence of the differential cross section on $p_{\rm{T}}^{miss}$.}
  \label{fig-miss_pt}
\end{figure}

Figure \ref{fig-pt_a} and  \ref{fig-miss_pt} show the differential cross sections as functions of
$p_{T}^{\gamma}$ and $p_{T}^{miss}$, respectively,  for the signal and backgrounds at the NLO level.
We can see that the differential cross sections of the
backgrounds decrease faster than that of the signal as the transverse momentum increases.
Thus, the ratio of signal and background can be improved if we set a larger $p_T$ cut.

\begin{figure}
  \includegraphics[width=0.6\linewidth]{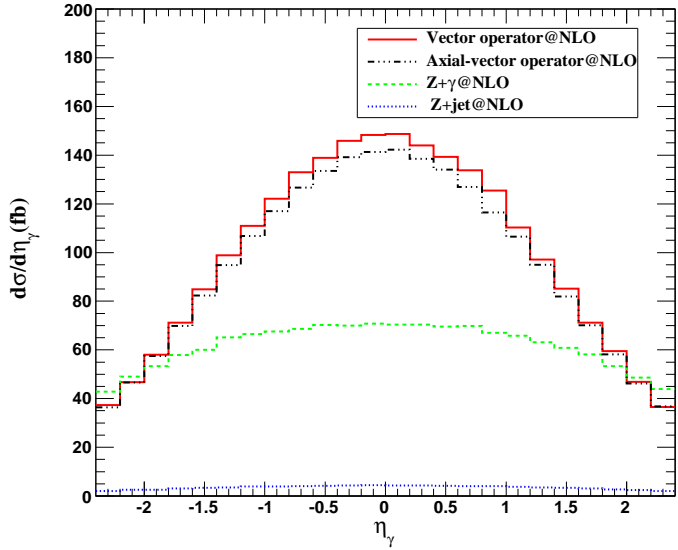}
  \caption{Dependence of the differential  cross section on $\eta^{\gamma}$.}
  \label{fig-eta}
\end{figure}

Figure \ref{fig-eta} shows the differential cross sections as a function
of  $\eta^{\rm{\gamma}}$  for the signal and the backgrounds
at the NLO level.
We see that the distribution of the signal is more concentrated in the central region
than the backgrounds.
These distributions give  some clues to suppress the backgrounds more efficiently at the LHC.

\begin{figure}
  \includegraphics[width=0.6\linewidth]{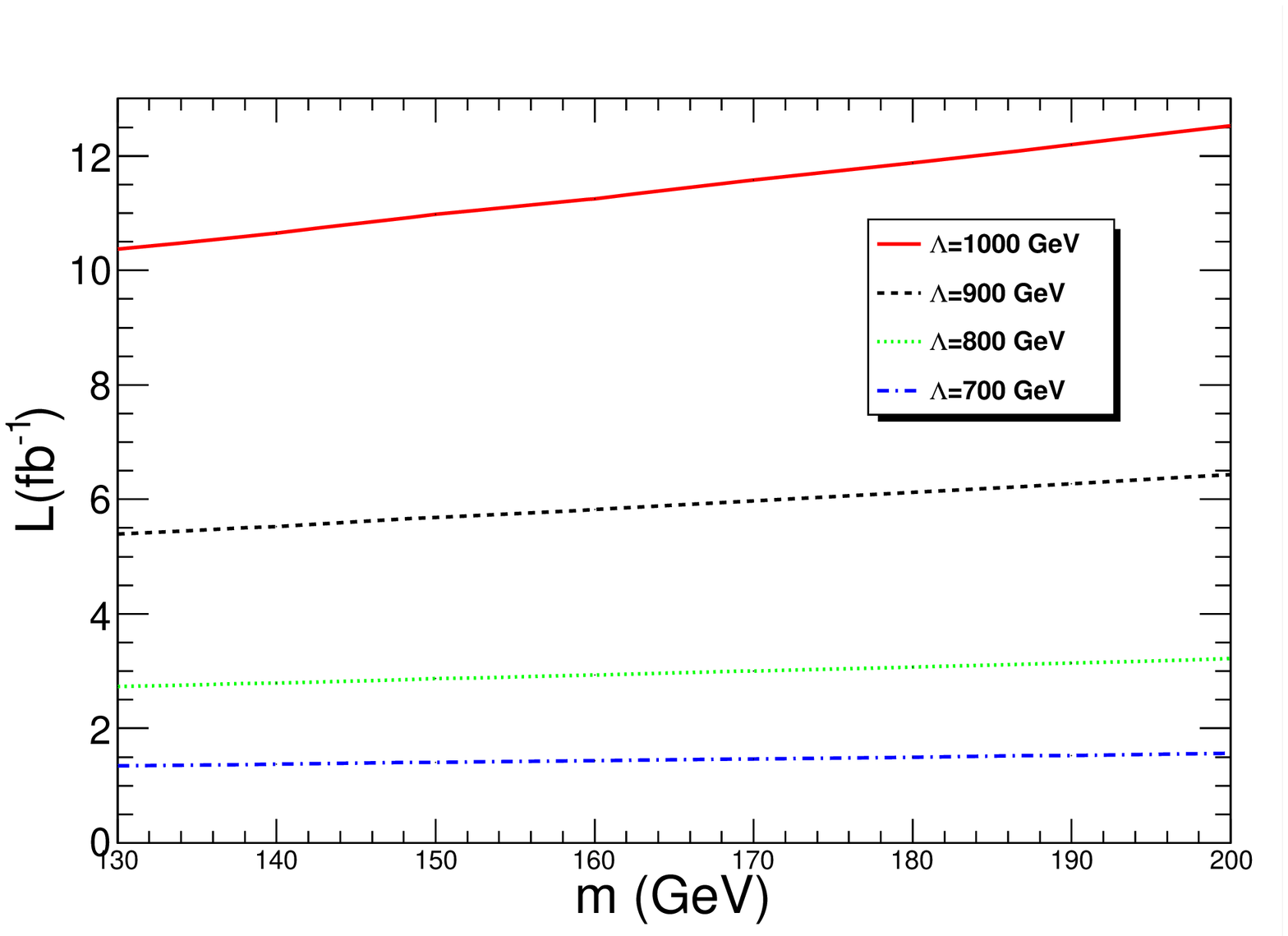}
  \caption{The integrated luminosity needed for a $5\sigma$ discovery as a function of  the DM mass  at the $14$ TeV LHC for the vector operator. We choose the cuts $p_T^{\gamma} > 300{\rm ~GeV}$ and $p_T^{miss} > 300{\rm ~GeV}$ due to the above analysis of Fig. \ref{fig-pt_a} and Fig. \ref{fig-miss_pt}.}
  \label{fig-discovery14}
\end{figure}

\begin{figure}
  \includegraphics[width=0.6\linewidth]{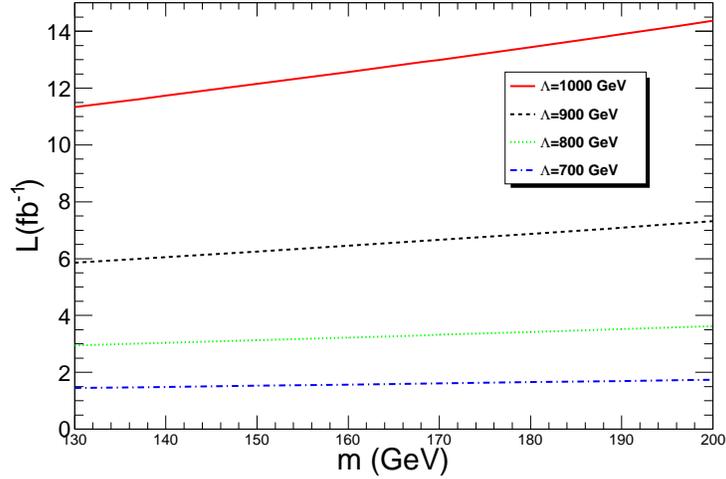}
  \caption{The integrated luminosity needed for a $5\sigma$ discovery as a function of  the DM mass  at the $14$ TeV LHC for the axial-vector operator. We choose the cuts $p_T^{\gamma} > 300{\rm ~GeV}$ and $p_T^{miss} > 300{\rm ~GeV}$ due to the above analysis of Figs. \ref{fig-pt_a} and  \ref{fig-miss_pt}.}
  \label{fig-avdiscovery14}
\end{figure}

\begin{figure}
  \includegraphics[width=0.6\linewidth]{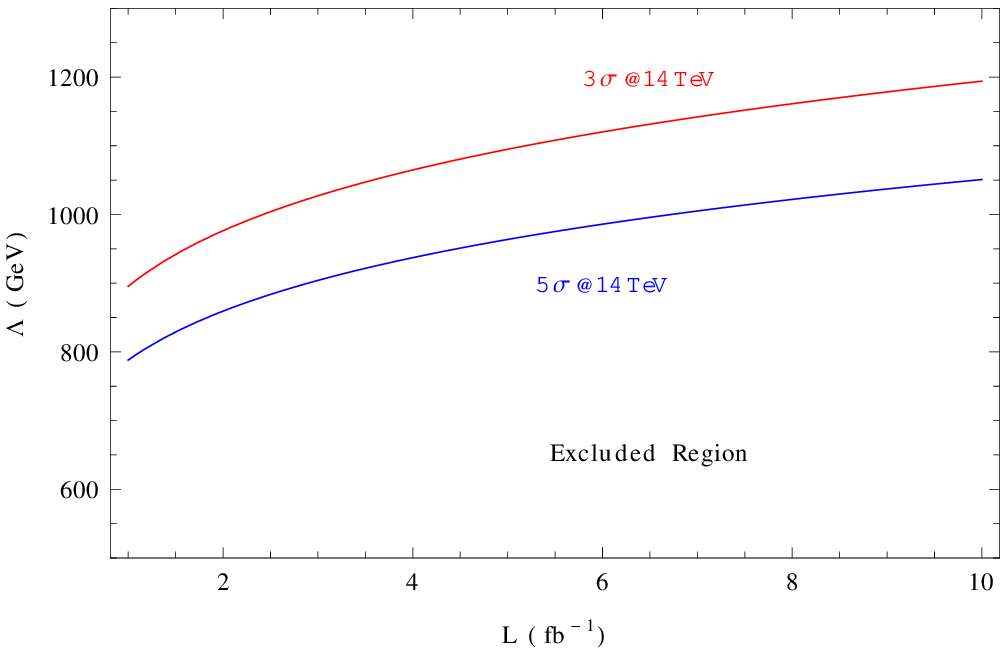}
  \caption{The limits of the NP scale for $3\sigma$ and $5\sigma$ exclusions at the $14$ TeV LHC, assuming $m=130$ ~GeV for the vector operator. We choose the cuts $p_T^{\gamma} > 300{\rm ~GeV}$ and $p_T^{miss} > 300{\rm ~GeV}$ due to the above analysis of Figs. \ref{fig-pt_a} and  \ref{fig-miss_pt}.}
  \label{fig-exclusion14}
\end{figure}

\begin{figure}[!h]
  \includegraphics[width=0.6\linewidth]{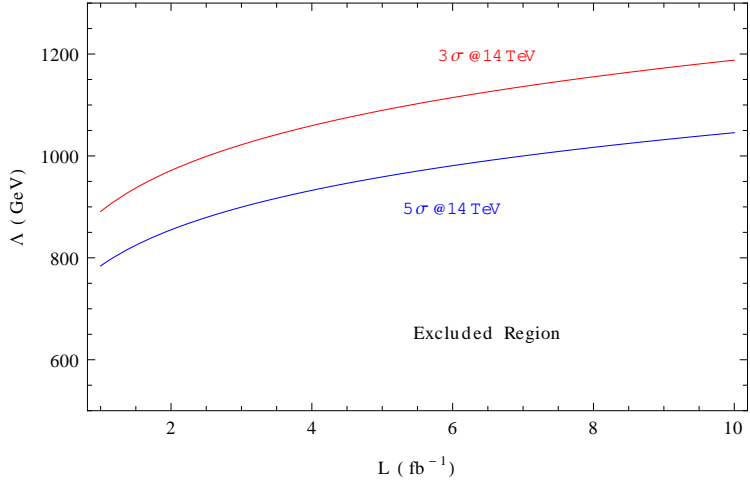}
  \caption{The limits of the NP scale for $3\sigma$ and $5\sigma$ exclusions at the $14$ TeV LHC, assuming $m=130$ ~GeV for the axial-vector operator. We choose the cuts $p_T^{\gamma} > 300{\rm ~GeV}$ and $p_T^{miss} > 300{\rm ~GeV}$ due to the above analysis of Figs. \ref{fig-pt_a} and \ref{fig-miss_pt}.}
  \label{fig-avexclusion14}
\end{figure}

Figure \ref{fig-discovery14} presents the integrated luminosity needed  to discover the signal
 at a $5\sigma$  level at the $14$ TeV  LHC.
We find that the needed integrated luminosity grows with the increasing
 of the NP scale, and  depends more strongly on the DM mass  for larger NP scale.
In particular, for $\Lambda=1000$ GeV and $m=200$ GeV, the needed integrated luminosity
is $12$ $\rm{fb}^{-1}$ at the $14$ TeV LHC. Figure \ref{fig-avdiscovery14} shows the results
for the axial-vector operator.
In Fig. \ref{fig-exclusion14}, we present the limits of the NP scale for
$3\sigma$  and $5\sigma$
exclusions at the $14$ TeV LHC, assuming $m=130$ ~GeV.
We see that the NP scale is constrained to be larger than $1200~{\rm GeV}$
if the $14$ TeV LHC  does not detect this signal after collecting an integrated
luminosity of $10~{\rm fb}^{-1}$.  Figure \ref{fig-avexclusion14} gives the results for the axial-vector
operator.

\section{Conclusion}\label{sec:conclusion}
We have investigated the signal of DM and photon associated production induced by
the vector and axial-vector operators at the LHC, including the QCD NLO effects.
We find that the QCD NLO corrections significantly reduce the dependence of the total
cross sections on the factorization and renormalization scales, and the QCD NLO corrections
are more significant for larger DM mass  for both the vector and axial-vector operators.
Using our NLO results, we improve the constraints on the NP scale from the  results of the
recent CMS experiment. Moreover, we calculate the dominant SM backgrounds at the NLO level,
and show the differential cross sections of both the signal and backgrounds as
functions of $p_{\rm{T}}^{\rm{\gamma}}$, $p_{\rm{T}}^{\rm{miss}}$ and $\eta^{\rm{\gamma}}$.
The character of these distributions can help to choose the kinematic cuts  in the experiments.
Finally, we show the  potential to discover the DM at the $14$ TeV LHC, and provide the exclusion
limits on the NP scale if this signal is not observed.

\acknowledgments
This work was supported by the National Natural
Science Foundation of China, under Grants
No. 11021092, No. 10975004 and No. 11135003.

\bibliography{avmyDDgamma}

\end{document}